\documentclass[preprint]{aastex}

\slugcomment{To appear in {\it The Astrophysical Journal Supplement Series}}

\begin{document}

\newfont{\fsym}{cmsy10 at 11 pt}

\title{The VLA 1.4~GHz Survey of the\\
Extended Chandra Deep Field South: Second Data Release}

\author{Neal A. Miller\altaffilmark{1}} 
\email{nmiller@astro.umd.edu}

\author{Margherita Bonzini\altaffilmark{2}}

\author{Edward B. Fomalont\altaffilmark{3}}

\author{Kenneth I. Kellermann\altaffilmark{3}}

\author{Vincenzo Mainieri\altaffilmark{2}}

\author{Paolo Padovani\altaffilmark{2}}

\author{Piero Rosati\altaffilmark{2}}

\author{Paolo Tozzi\altaffilmark{4}}

\author{Shaji Vattakunnel\altaffilmark{4}}

\altaffiltext{1}{Department of Astronomy, University of Maryland, College Park, MD 20742}
\altaffiltext{2}{European Southern Observatory, Karl-Schwarzschild-Strasse 2, D-85748 Garching bei M\"unchen, Germany}
\altaffiltext{3}{National Radio Astronomy Observatory, 520 Edgemont Road, Charlottesville, VA \ 22903}
\altaffiltext{4}{INAF Osservatorio Astronomico di Trieste, via G.B. Tiepolo 11, I-34131, Trieste, Italy}

\begin{abstract} 
Deep radio observations at 1.4~GHz for the Extended Chandra Deep Field South were performed in June through September of 2007 and presented in a first data release (Miller et al. 2008). The survey was made using six separate pointings of the Very Large Array (VLA) with over 40 hours of observation per pointing. In the current paper, we improve on the data reduction to produce a second data release (DR2) mosaic image. This DR2 image covers an area of about a third of a square degree and reaches a best rms sensitivity of 6 $\mu$Jy and has a typical sensitivity of 7.4 $\mu$Jy per 2\farcs8 by 1\farcs6 beam. We also present a more comprehensive catalog, including sources down to peak flux densities of five or more times the local rms noise along with information on source sizes and relevant pointing data. We discuss in some detail the consideration of whether sources are resolved under the complication of a radio image created as a mosaic of separate pointings each suffering some degree of bandwidth smearing, and the accurate evaluation of the flux densities of such sources. Finally, the radio morphologies and optical/near-IR counterpart identifications (Bonzini et al. 2012) are used to identify 17 likely multiple-component sources and arrive at a catalog of 883 radio sources, which is roughly double the number of sources contained in the first data release.
\end{abstract}

\section{Introduction}\label{sec-intro}

The {\it Chandra} Deep Field South (CDF-S) continues to be one of the most important deep fields for multiwavelength investigation of the cosmological evolution of galaxies. Originally devoted 1 Msec of {\it Chandra} integration \citep{giacconi2002}, subsequent observations have pushed ever deeper in X-ray flux by increasing the net exposure first to 2 Msec \citep{luo2008} and recently to 4 Msec \citep{xue2011}. This makes the CDF-S the most sensitive X-ray view of the universe, reaching sources with full-band (0.5 - 8.0 keV) fluxes down to $\approx 3.2 \times 10^{-17}$ erg~s$^{-1}$~cm$^{-2}$. The area covered by relatively deep X-ray observations has also increased over the initial survey, with an additional one Msec divided among four {\it Chandra} pointings to produce the Extended CDF-S \citep[E-CDF-S;][]{lehmer2005}. Likewise, {\it XMM-Newton} has observed the field in increments over the past nine years and has now accumulated over 3 Msec of integration in a region that covers much of the E-CDF-S. These data will provide quality X-ray spectra of detected sources \citep[e.g.,][]{comastri2011}. 

In addition to X-ray coverage, {\it Hubble} and {\it Spitzer} have produced deep near-ultraviolet through infrared data for the field. The deepest images ever obtained with {\it Hubble} are located in the CDF-S, with the {\it Hubble} Ultra Deep Field using the Advanced Camera for Surveys (ACS) to obtain deep $F435W$ ($\sim B$), $F606W$ ($\sim V$), $F775W$ ($\sim i$), and $F850LP$ ($\sim z$) images \citep{beckwith2006} and more recently the Wide Field Camera 3 (WFC3) instrument exploring the near-IR $F105W$ ($Y$), $F125W$ ($\sim J$), and $F160W$ ($\sim H$) with a 192-orbit ``Treasury'' program (PI G.\ Illingworth). The Great Observatories Origins Deep Survey \citep[GOODS;][]{giavalisco2004} targets the CDF-S as one of its two fields, using both {\it Hubble} and {\it Spitzer} images to study galaxy evolution across cosmic time. Analogous to the widening of the CDF-S to the E-CDF-S in X-ray surveys, programs have expanded the covered area using {\it Hubble} \citep[Galaxy Evolution from Morphologies and SEDs, GEMS;][]{rix2004} and {\it Spitzer} \citep[the {\it Spitzer} IRAC/MUSYC Public Legacy Survey, SIMPLE, at near-IR wavelengths and the Far-Infrared Deep Extragalactic Legacy survey, FIDEL, in mid-IR;][]{damen2011,magnelli2011}. Most recently, the ``Cosmic Assembly Near-IR Deep Extragalactic Legacy Survey'' (CANDELS) is capitalizing on {\it Hubble}'s WFC3 camera to provide deep near-IR imaging of several fields including the CDF-S \citep{grogin2011,koekemoer2011}. In terms of allocated orbits, CANDELS is the largest project in the history of {\it Hubble} and its three near-IR filters will identify and characterize galaxies over the approximate redshift range $1.5 \leq z \leq 8$. {\it Spitzer} observations continue with warm mission {\it Spitzer} activity \citep[e.g., the ``{\it Spitzer} Extragalactic Representative Volume Survey,'' or SERVS\footnote{See also the {\it Spitzer} Extended Deep Survey, SEDS ({\ttfamily http://www.cfa.harvard.edu/SEDS/index.html}).};][]{mauduit2012}. Finally, {\it Herschel} has pushed the deep IR data into the far-IR portion of the spectrum with the GOODS-{\it Herschel} program \citep{elbaz2011} and the {\it Herschel} Multitiered Extragalactic Survey \citep[``HerMES;''][]{oliver2012}.

Ground-based observatories have also supplied much ancillary data at optical and near-IR wavelengths. These programs include $J~H~K_s$ imaging \citep{retzlaff2010} with the Infrared Spectrometer And Array Camera (ISAAC) of the European Southern Observatory's Very Large Telescope (VLT) and $U~B~V~R~I$ imaging from the Garching-Bonn Deep Survey \citep{hildebrandt2006} using the Wide Field Imager (WFI) at the 2.2m La~Silla telescope. The E-CDF-S is also one of four fields covered by the Multiwavelength Survey by Yale-Chile \citep[MUSYC;][]{gawiser2006}, with this program using the Cerro Tololo Inter-American Observatory (CTIO) 4~m telescope and the MOSAIC~II and ISPI instruments to obtain deep coverage of the E-CDF-S region at $U~B~V~R~I~z^\prime~J~H~K_s$. The MUSYC project has also acquired medium-band imaging in 18 filters using the Subaru Telescope and used these along with the broad-band and {\it Spitzer} IRAC photometry to produce accurate photometric redshifts for tens of thousands of galaxies \citep{cardamone2010}. In addition to photometric redshifts, there are a host of spectroscopic programs providing critical spectroscopic redshifts and source classifications. These have often been associated with the GOODS program and capitalized on the VLT with the FORS2 and VIMOS instruments \citep[][and associated papers]{vanzella2008,balestra2010}. Many other spectroscopic campaigns have also targeted the field, including specific targeting of radio-selected sources \citep[e.g.,][]{mao2012} and recent efforts to expand the area from the well-sampled GOODS region (essentially the CDF-S) to the full 30\arcmin{} by 30\arcmin{} E-CDF-S \citep[e.g.,][using the Inamori-Magellan Areal Camera and Spectrograph, IMACS, on the Magellan-Baade telescope]{cooper2012}.

The field has also received extensive attention at radio wavelengths. \citet{norris2006} used the Australia Telescope Compact Array (ATCA) at 1.4~GHz to observe a large (3.7 deg$^2$) area, with the E-CDF-S region reaching down to an rms sensitivity of $\sim15\mu$Jy per $17^{\prime\prime} \times 7^{\prime\prime}$ beam. \citet[][hereafter K08]{kellermann2008} used the National Radio Astronomy Observatory (NRAO) Very Large Array (VLA) to obtain deeper and higher resolution 1.4~GHz data ($8.5\mu$Jy rms noise per $3\farcs5 \times 3\farcs5$ beam) in a field centered on the CDF-S. These latter data were exploited in a sequence of papers on optical counterparts \citep{mainieri2008}, radio-X-ray properties \citep{tozzi2009}, and source populations and evolution \citep{padovani2009,padovani2011}. In addition to 1.4~GHz, VLA observations at 5~GHz are presented in K08 with an rms sensitivity of $8.5\mu$Jy rms for a $3.5^{\prime\prime} \times 3.5^{\prime\prime}$ beam and \citet{huynh2012} present an ATCA 5.5~GHz survey of the E-CDF-S with a $\sim12\mu$Jy rms per $4.9^{\prime\prime} \times 2.0^{\prime\prime}$ beam. At longer wavelengths, \citet{ivison2010} included Giant Metre-Wave Radio Telescope data at 610~MHz ($\sim40\mu$Jy rms noise for a $6.5^{\prime\prime} \times 5.4^{\prime\prime}$ beam) in their investigation of evolution in the FIR-radio correlation.

In June through September 2007, we observed the E-CDF-S with the National Radio Astronomy Observatory (NRAO)\footnote{The National Radio Astronomy Observatory is a facility of the National Science Foundation operated under cooperative agreement by Associated Universities, Inc.} Very Large Array (VLA) under program code AM889. The observations were performed at a frequency of 1.4~GHz and consisted of over 250 hours of time (about 690~ksec time on source) spread across six separate pointings. The survey description, strategy, and initial image and catalog were presented in \citet{miller2008}. The rapid turnaround time between data collection and publication was motivated by the strong community interest in the field, which has been justified by the number of studies which have relied upon the data to set or confirm the astrometry for observations made at other wavelengths \citep[e.g.,][]{truch2009,weiss2009,scott2010,xue2011}, identify counterparts to sub-millimeter surveys \citep{coppin2009,dye2009,biggs2011,yun2011}, extend the far-infrared/radio correlation out to cosmological redshifts \citep{ivison2010,bourne2011,mao2011}, and identify and study star-forming galaxies and active galactic nuclei \citep{moncelsi2011,vattakunnel2011,fiore2011,danielson2012}. Naturally, this rapid release of the calibrated radio images necessitated some minor compromises. In regards to the imaging of the data, these amounted to postponing some time- and computational-intensive techniques that produce slight improvements to the depth and cosmetics of the resulting images. Similarly, the initial catalog was conservative in only going to a 7$\sigma$ point-source detection limit and providing little detail about source morphology. In the current paper, we present the second data release (DR2) associated with this program. It incorporates these finer imaging techniques yielding a typical reduction in the rms noise of about 0.5 $\mu$Jy across the full E-CDF-S area, and this improvement plus moving to a 5$\sigma$ detection threshold produces a deeper and more comprehensive but still highly-reliable source catalog. 

We describe the details associated with the improvements to the imaging in Section \ref{sec-image}. In Section \ref{sec-cat} we discuss the detection and characterization of sources applied in the construction of a source catalog, and briefly discuss the catalog and future directions in Section \ref{sec-sum}.

\section{Improved Imaging}\label{sec-image}

\subsection{Background}

To fix our terminology for the ensuing discussion, we provide a brief summary of the observations. In order to cover the full E-CDF-S area at near-uniform sensitivity, we pointed the VLA at six separate coordinate locations arranged in a hexagonal grid around the adopted center of the CDF-S, (J2000) $03^h32^m28.0^s$ $-27^\circ48^\prime30.00^{\prime\prime}$. We refer to these as ``pointings'' and often reference them with a numerical designation from 1 to 6 starting due east of the center coordinate and proceeding clockwise; Table \ref{tbl-ptgs} includes the numerical designations and coordinate centers. Our observations were spread over many days on account of the low declination of the field and typically amounted to five hours of time per calendar date. Thus, the full program consists of about fifty of these ``tracks.'' We opted to observe a single pointing on any given calendar date, meaning that each pointing was observed on at least eight separate dates. While this was done largely for efficiency and convenience, it also allows for a deep investigation of possible radio transient populations (Frail et al., in preparation). We refer those interested in the survey strategy and design to \citet[][hereafter M08]{miller2008}. 
 
\subsection{General Procedure}

At the time of the observations the VLA was undergoing improvements paving the way toward full ``Expanded'' VLA \citep{perley2011} operations, a now complete process that includes the official renaming as the Karl G.\ Jansky Very Large Array. In general, these modifications to the array had little effect, and the procedures for calibrating and imaging the data followed the standard prescriptions for deep, multi-channel continuum VLA observations at 1.4~GHz using NRAO's Astronomical Image Processing System (AIPS). The most notable exception was the retirement of the original VLA control computers on 2007 June 27, which led to slight errors in how the $(u,v,w)$ coordinates were written at various times during the range of dates covering our observations. For the first data release, we relied upon the AIPS task UVFIX to recalculate the $(u,v,w)$ data based on the antenna positions which are included among the data tables associated with each observation. Several of the observations were also afflicted by a glitch which reversed the channel indexing for brief periods of time, and we fixed this error using a sequence of existing AIPS procedures. Finally, in a few instances the most-recently retrofitted EVLA antenna had incorrect system temperature values which produced spuriously large weights that we had to manually adjust down to more representative values. 

Each of these problems has now been corrected in the NRAO Data Archive, and in this second pass at data calibration and imaging we essentially started ``from scratch.'' We re-obtained the raw $(u,v,w)$ data from the Archive, and followed the same general procedure of data editing, calibration, and imaging as was described in M08. One minor improvement was the inclusion of a very small amount of extra data relative to that previous reduction. As VLA antennas are retrofit to become EVLA antennas, they are placed on a ``master pad'' for testing. There were some times during the course of our observations when such antennas were included in the array and we were able to calibrate such data and include it in our current imaging. Although we mainly started from scratch and paralleled the previous reduction procedure, we did have one significant saving: the presence of first data release deep images associated with each pointing. These were used to self-calibrate the $(u,v,w)$ data after the initial data edits and amplitude and phase calibration. 

The basic sequence of initial steps is identical to that presented in M08. For each observing track, we inspected and edited the calibrator $(u,v,w)$ data (3C48 and J0340-213). These data were then used to establish the bandpass calibration (using 3C48), flux density scale (from 3C48, with J0340-213 bootstrapped to 3C48), and phase calibration (using J0340-213) for the target data. After application of the calibrations, the $(u,v,w)$ data for the target were inspected and edited to remove obvious errors. The resulting $(u,v,w)$ data were then self-calibrated on phase using the deep individual pointing images created for the first data release. We proceeded to then image the data, subtract the determined clean components, edit the resulting source-removed $(u,v,w)$ file to flag obvious interference and data errors, and then return the clean components. Once all the data associated with an individual pointing were thus calibrated and edited, we combined them into single, deep $(u,v,w)$ data set associated with that pointing. 

These combined single-pointing datasets were then imaged and subjected to further self-calibration and editing. First, we applied a strong taper to the $(u,v,w)$ data and generated a wide-field, low resolution map. Over this image we laid the grid of 127 facets arranged in a ``flys-eye'' pattern that would be used in our imaging, with these smaller facets allowing for correction of image distortions caused by sky curvature. Each of these facets was $512 \times 512$ 0\farcs5 pixels, and the flys-eye thus covered about 52\arcmin{} in diameter -- not quite out to the first null of the VLA primary beam at 1.4~GHz. We were thus able to search the wide-field image for additional faint sources outside of our principal imaged area, including a small number within the primary beam and a larger number within the sidelobes. While these same basic steps were performed previously on individual tracks of data, repeating them on the deeper combined datasets warranted an increase in the radial coverage of individual facets (previously 91 were used) and revealed additional faint sources outside the flys-eye coverage. We identified about 30 such additional small fields to image for each pointing. Finally, we searched within the main 127-facet coverage for particularly bright sources (those with flux densities uncorrected for primary beam response of a few mJy and higher). This identified between seven and 10 additional small facets for imaging, and thus in total each pointing was imaged as the combination of between about 160 and 175 facets. 

Successive rounds of imaging and self-calibration were then performed, with each newer set of images used as the source model for self-calibration in both amplitude and phase. Our final steps of self-calibration on the full data for each pointing used the AIPS task {\scshape peelr}, which ``peels'' off individual bright sources to improve the overall calibration. These sources are returned to the data once all corrections are made. Between one and four bright sources per field were subjected to {\scshape peelr}, with these sources typically having an apparent flux density (uncorrected for primary beam attenuation) of about 7.5 mJy and greater.

The final imaging was done in segments of the full $(u,v,w)$ data. This provides a fine correction to several effects, each of which relates to the shape of the primary beam and how it can differ slightly for different portions of the data. These minor differences cause the response to sources to vary over the course of an observing track. First, the VLA observed in a pair of frequency bands (intermediate frequencies, or IFs) that bracket 1.4~GHz. These separate IFs imply slightly different resolutions. Second, the feeds for the right and left circular polarization are not coincident and thus have slightly different pointing centers. Third, the power pattern of the VLA primary beam is not perfectly circular and over the course of observations as the hour angles of sources change their responses vary. We consequently separated the data into 12 segments, consisting of three ranges in hour angle and separately for each IF and polarization. As before, we imaged each segment of the data and performed a round of self-calibration on that segment before producing the final image for that segment. The images corresponding to the 12 segments for each pointing were then combined using variance weighting based on their individual rms sensitivities. Relative to the first data release, the final images for the six pointings had net improvements in sensitivity ranging from about $0.1\mu$Jy to $0.7\mu$Jy. Table \ref{tbl-ptgs} includes the achieved rms sensitivity per pointing.

\subsection{Summary of Final Data and Images}\label{sec-image:sum}

\begin{figure}
\figurenum{1}
\epsscale{0.9}
\plotone{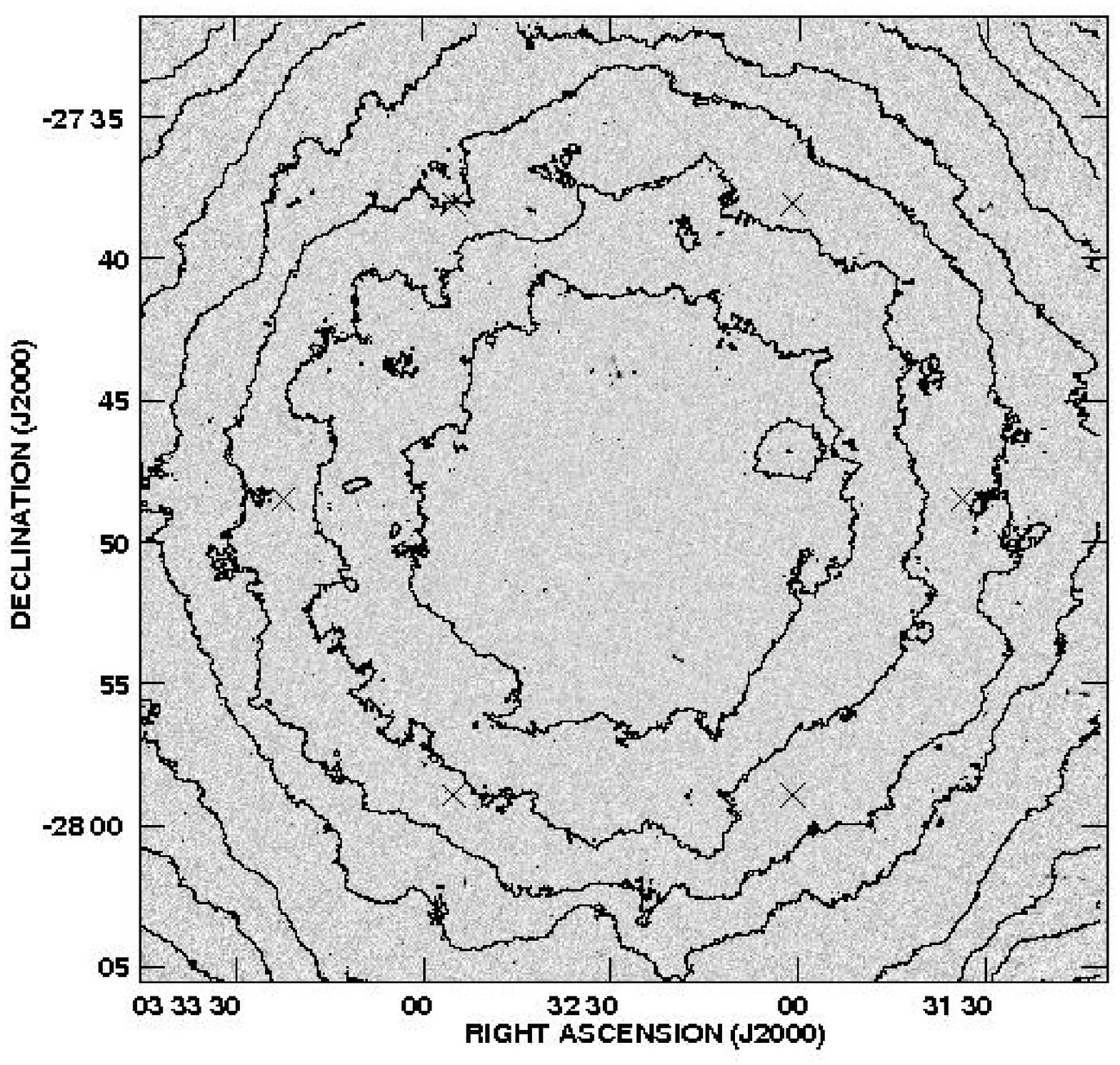}
\caption{Greyscale depiction of the DR2 mosaic image, with overlaid contours of constant RMS noise. From the center, these contours represent 6.5 $\mu$Jy, 7.0 $\mu$Jy, 7.5 $\mu$Jy, 8.0 $\mu$Jy, 9 $\mu$Jy, 10 $\mu$Jy, 11 $\mu$Jy, and 12 $\mu$Jy per beam. The six pointing centers for the observations (Table \ref{tbl-ptgs}) are indicated by crosses.\label{fig-image}}
\end{figure}

The images corresponding to the six individual pointings were then combined to form the final mosaic image. The mosaic step corrects for the power pattern of the primary beam, and data out to the radial distance where the power pattern is 33\% of that at the pointing center were included (i.e., just under 20\arcmin{} in radius). This choice of radial cutoff is somewhat arbitrary but was motivated by the consideration that data this distant from a pointing center are de-weighted by about an order of magnitude in creation of the mosaic image. It is also consistent with that used in other radio surveys \citep[e.g.,][]{huynh2005,schinnerer2010}. The contribution at each point in the output mosaic is weighted by the inverse of the power pattern squared at that point. We also included a single track of data from early in the observing campaign where we had inadvertantly shifted the pointing center by 1\arcmin . The area covered by the final mosaic is $\sim$34\arcmin{} $\times \sim$34\arcmin{} (0.324 square degrees), being $4096 \times 4096$ 0\farcs5 square pixels. This is the main image associated with this second data release and is shown in Figure \ref{fig-image}. It will be used in subsequent characterization and compilation of a source catalog. We note that the actual coverage of the individual pointings extends well beyond the boundaries of this mosaic image but experiences the drop-off in sensitivity associated with the power pattern of the VLA primary beam. Users interested in sources that fall within the coverage of such pointings but outside the final mosaic image are directed to the individual pointing images and a larger mosaic image that incorporates all of the data.

The sensitivity and coverage thereof was evaluated via the construction of an rms sensitivity map. To construct this map, we took the final mosaic image and removed all sources with peak flux densities greater than 150 $\mu$Jy, chosen as approximately 20$\sigma$ point source detections. The rms noise at each pixel in the mosaic map was then determined from this bright-source-removed map, based on the value calculated within a 135\arcsec{}-diameter circle centered on that pixel. This ``background mesh'' size is considerably larger than some other radio surveys which suggest that sizes on the order of just ten beam widths across are sufficient \citep[e.g.,]{schinnerer2010,huynh2012}. We found that, given the resolution of our survey and the existence of extended sources that approached the size of these smaller background mesh apertures, their usage produced inaccurate small-scale variations in rms sensitivity maps. The larger size that we used was also consistent with what was used in the first data release, although in that release the background mesh was square but of the same equivalent area. The evaluation of the rms noise in constructing the rms sensitivity map was iterative, with points differing by more than three times the calculated rms noise removed prior to re-calculation until convergence was achieved. The final sensitivity map, hereafter called the RMS map, is relatively smooth as seen by the overlaid contours of constant rms included in Figure \ref{fig-image}.

Figure \ref{fig-rms} depicts the area covered at a given sensitivity or better based on this RMS map. The most sensitive regions of the mosaic image have an rms noise of 6 $\mu$Jy per beam, and half of the full mosaic image has an rms noise of 7.4 $\mu$Jy per beam or better. This represents an improvement of about 0.5 $\mu$Jy over the first data release mosaic image. The regions at the center of the mosaic where all six individual pointings can contribute are the most sensitive, and these correspond to the CDF-S coverage. The CDF-S 4~Msec X-ray data correspond to 54 separate {\em Chandra} images collected using ACIS-I and thus each having a field size of about 16\farcm9 by 16\farcm9. The exposure-weighted center of these data is (J2000) $03^h32^m28.06^s$ $-27^\circ48^\prime26.4^{\prime\prime}$ \citep{xue2011}, and for the range of roll angles and slight shifts in aim point of the separate exposures it is reasonable to assume approximately uniform coverage out to about 7$^\prime$ in radius from this central point. Within this restricted area of the deepest X-ray coverage (0.043 square degrees, or 13\% of the mosaic image) the radio mosaic image has a typical rms noise of 6.3 $\mu$Jy and is never worse than 6.7 $\mu$Jy. This is depicted with the dotted line in Figure \ref{fig-rms}. We also define a subset of the full mosaic over which there are enough ancillary multiwavelength data to provide counterpart identification and spectroscopic redshifts or reasonable photometric redshifts (Padovani et al., in preparation; this region amounts to 0.282 square degrees or 87\% of the mosaic image). The rms sensitivity as a function of area for this restricted region is depicted by the dashed line in Figure \ref{fig-rms}, and is used in correcting the source counts for investigation of contributions by source type.


Our data release consists of the six $(u,v,w)$ data sets, nine images, and one catalog. The $(u,v,w)$ data are the final combined and edited collections for each of the six pointings, and allow interested users to perform imaging with their own choices of weighting parameters. The nine images correspond to the final mosaic image and its associated RMS map, the final images corresponding to each of the six individual pointings (each with an rms near 10 $\mu$Jy at the field center), and a large mosaic map made from these six individual pointing images and representing the full imaging coverage (i.e., out past the coverage of the main final mosaic image). The catalog is based entirely on the final mosaic map, and will be discussed now in greater detail.

\section{Source Catalog}\label{sec-cat}

\subsection{Source Detection}

The basis for the detection of sources was a signal-to-noise ratio ($SNR$) map, constructed by dividing the final mosaic image by its corresponding RMS map. The AIPS task {\scshape sad} was used to pick out all sources in this $SNR$ map with peaks above above 3, and thus three times the local noise. We note that the effect of using an RMS map generated with a larger background mesh size is to slightly increase the number of sources above the detection threshold by smoothing over small-scale variation produced by the presence of bright sources and the more rapid fall-off in sensitivity at the edges of the mosaic. {\scshape sad} fits Gaussians to the sources above the 3$\sigma$ threshold and subtracts them from the $SNR$ map to produce a residual image which may then be inspected to find sources that were either missed or poorly fit. These include sources with extended radio morphologies as well as blends of multiple and possibly independent radio sources. The residual image was inspected by eye with the help of overlaid contours to reveal such missed or poorly fit sources (indicated by large positive or negative values) and these were manually added to the list for subsequent analysis.

We then proceeded to evaluate all sources with peaks above 4$\sigma$ identified in the $SNR$ image by either {\scshape sad} or our manual inspection. Our initial choice of 3$\sigma$ was based on prior knowledge that sometimes single faint sources might be incorrectly split into a pair of fainter sources, and we used the {\scshape sad} list to recover these objects as multiple $<4\sigma$ objects at nearly identical position. In this round of source cataloging, we used the AIPS task {\scshape jmfit} to fit Gaussians to the sources on the final mosaic image in order to evaluate source peak and integrated flux densities. {\scshape jmfit} was provided with a guess for the peak flux density and its position (based on manual inspection of the image), but no corrections for primary beam attenuation or bandwidth smearing were applied. A correction for the former was explicitly applied in construction of the mosaic image, while the latter is complicated by the lack of a single pointing center for the image. The output parameters of {\scshape jmfit} included the source coordinates, major and minor axis size and position angle, and the peak and integrated flux density. We used the RMS map to determine the local noise at the coordinates of the fitted source, and accepted those sources with peak flux density greater than five times their local noise value for inclusion in the final catalog. For those sources poorly fit by Gaussians in the initial investigation using the $SNR$ map, we directed {\scshape jmfit} to fit multiple components where applicable and performed aperture photometry on the more extended sources using the {\scshape tvstat} task. 

For some work correlating the radio catalog with other wavelengths, it might seem of value to release the initial automated catalog that extended to down 3$\sigma$. However, we deem the danger of misinterpretation of such low-significance sources outweighs the possible utility of such a catalog. In addition to some issues already discussed (i.e., false decomposition of single faint sources), we note here that if the mosaic image were truly described by Gaussian noise there would be over 1,100 noise peaks greater than 3$\sigma$ across its large area. This number becomes reasonable at 4$\sigma$, where about 26 noise peaks would be expected. Using the traditional $5\sigma$ detection threshold the number of anticipated false sources has dropped to less than one across the full area of the final mosaic. The true noise distribution is not perfectly Gaussian as a result of the presence of real sources and their sidelobes, so these numbers may be considered lower limits to the estimate of the number of false sources at these thresholds. Finally, we note that the availability of the actual mosaic image allows users to directly assess possible low-significance radio sources at their positions of interest.

\subsection{Source Morphology}\label{sec-morph}

Source morphology is an important aspect of source catalogs, often providing a coarse way to discriminate between emission mechanisms at appropriate resolutions. Radio emission associated with star formation is generally extended across galaxy disks, and thus resolved at arcsecond scales even for sources at high redshifts \citep{muxlow2005}. Evaluation of the true flux density of a source is also often dependent on source morphology, with compact sources at lower signal-to-noise ratio better described by their peak flux density than by their integrated flux density \citep[e.g., see ][]{owen2008}. 

Unfortunately, the careful consideration of source morphology is complicated by the mosaic nature of the final image. Every location in the output mosaic is the combination of the contributions of up to six separate pointings, and thus up to six separate corrections each dependent on differing values of angle and radial distance. If these corrections do not alter the apparent morphology of a source but only its apparent flux density, they are easily applied. This is the case for primary beam attenuation, and the simplification that the power pattern is radially symmetric is applied and accounted in the creation of the mosaic map. However, other effects such as bandwidth smearing are not so easily handled. A given location in the output mosaic is the combination of up to six separate images each of which suffers some degree of smearing in the radial direction relative to its relevant pointing center, and while corrections for this effect may be made on sources extracted from single-pointing data they are not so easily achieved when multiple pointings are combined. Bandwidth smearing does not change the integrated flux density in the limit of a noiseless image, and in M08 we confirmed that for the E-CDF-S radio survey the standard Gaussian fitting for source integrated flux densities provided consistent results with aperture photometry. However, the combination of up to six separate bandwidth smearing corrections does affect source morphology and the determination of whether a source is resolved.

In light of this complication, we performed multiple tests of source morphology that are encapsulated in the source catalog. The simplest of these is the direct fit to the mosaic image, which produces a major axis, minor axis, and position angle of the Gaussian which best fits the source in the mosaic image. In addition to ignoring effects such as bandwidth smearing, these results are also still the convolution of the primary beam and the intrinsic morphology of the source. As such, they are not tremendously useful except in the very general sense of providing a quick handle on which sources are clearly resolved (i.e., those with axial sizes well above the 2\farcs8 by 1\farcs6 beam size). Similarly, sources in the mosaic image which were poorly fit were identified in the residual maps produced as the difference between the mosaic image and the best Gaussian fits to sources. The presence of large residuals is strongly suggestive of an extended source, and the integrated flux density for these sources was evaluated using manually-sized aperture photometry with the task {\scshape tvstat}.

For single pointing data we can incorporate the effect of bandwidth smearing into the source fitting. The distances between each source and the six separate pointing centers was determined, and when such distances were less than the 33\% power point of the primary beam these pointings were noted. This provides a listing of the pointings which directly contributed to the mosaic image at the location of each source. We then determined the rms sensitivity of each contributing pointing at the coordinates of each source in order to determine which individual pointing provided the best signal-to-noise for source measurement. This is usually the pointing nearest to the source, but slight differences in achieved depth per pointing did occasionally mean a more distant pointing provided a marginally lower rms sensitivity. We then evaluated the Gaussian fit to each source using just this ``best'' single pointing, incorporating the effects of both primary beam attenuation and bandwidth smearing. The parameters of these fits, deconvolved to remove the primary beam geometry, are included in the source catalog. The major and minor axis parameters are provided as $\pm1\sigma$ ranges, so that sources with fitted axes differing from zero may be identified and thus provide an evaluation that a source is resolved (i.e., the minimum size for its major axis is greater than zero). This represents the first of our two tests on whether a source is resolved.

\begin{figure}
\figurenum{3}
\epsscale{0.9}
\plotone{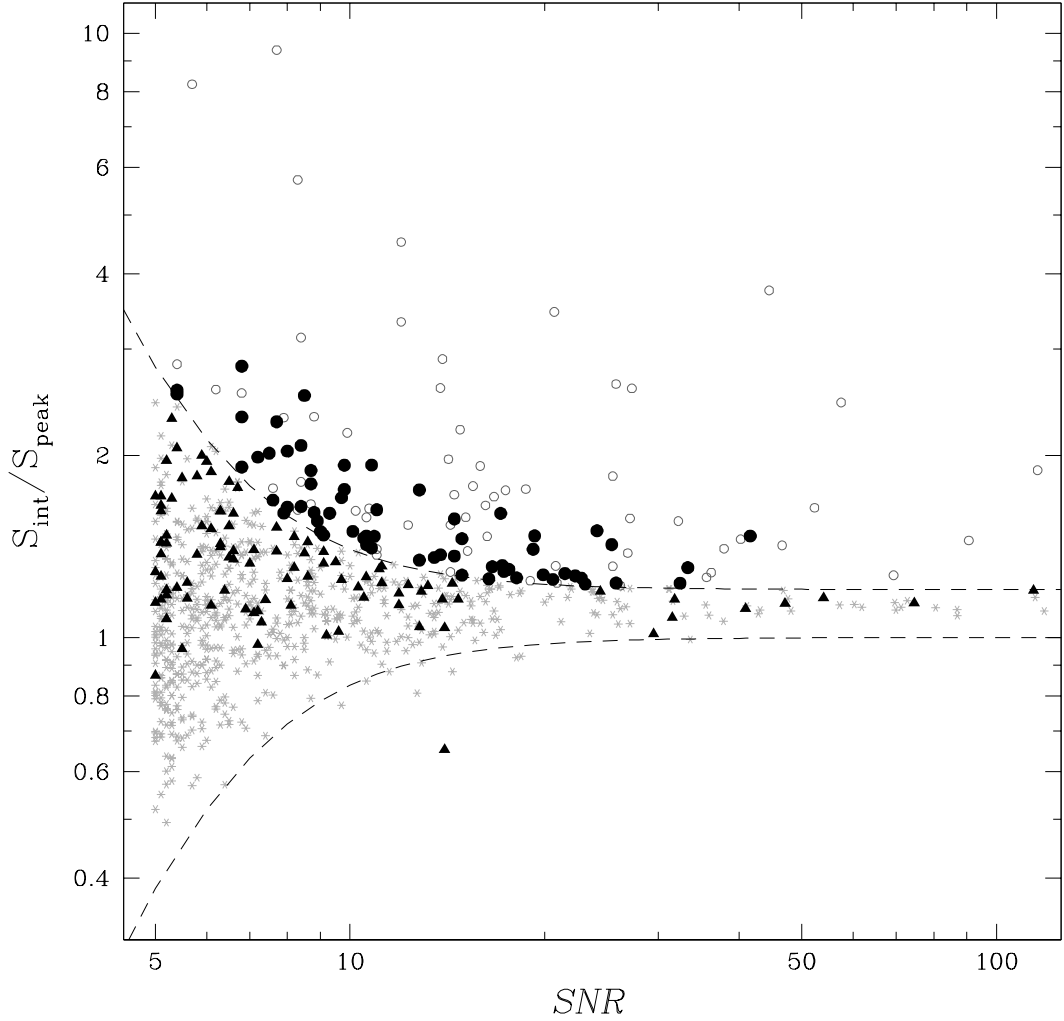}
\caption{Analysis of whether sources are resolved, based on $SNR$ and the ratio of integrated to peak flux densities from the mosaic image. The dashed lines show the envelope within which sources are assumed to be unresolved, with sources above the upper dashed line being resolved according to this approach. Grey asterisks are sources which are unresolved by both this method and investigation of their fitted axes in single-pointing data (including corrections for primary beam response and bandwidth smearing), and open circles are objects for which each method indicates a resolved source. Sources whose classification differs between the two methods are shown by filled black points, with filled circles being resolved according to this method but not by the Gaussian fit to their individual pointing data and filled triangles being the opposite.\label{fig-resck}}
\end{figure}

A second technique to evaluate whether a source is resolved involves comparison of the fitted integrated and peak flux densities. A comparison of the ratio of integrated to peak flux density as a function of signal-to-noise ratio (defined as the peak flux density divided by the local rms sensitivity) is an empirical estimate of the errors in source fitting, and can be used to identify resolved sources as those for which the integrated flux density is clearly in excess of the peak flux density after accounting for the fitting errors \citep[e.g.,][]{huynh2005}. Although this method does not provide direct indicators of source sizes and morphologies, it is quite helpful in the determination of the better estimate of a source's true flux density. In Figure \ref{fig-resck} we plot the ratio of the integrated to peak flux densities as a function of signal-to-noise based on the Gaussian fits to sources in the mosaic image. Situations where the integrated flux density is less than the peak flux density are clearly the result of errors inherent to source fitting, and we can determine a function which bounds these cases. Under the assumption that it is equally likely that such errors can result in fits for which the integrated flux density is greater than the peak flux density, we can mirror this function above the line where the peak flux density equals the integrated flux density and thus define an envelope which contains sources that are consistent with being unresolved. Sources for which the ratio of integrated to peak flux density is greater than this envelope are those that are likely to be resolved. 

We follow the functional form presented in \citet{huynh2005} to fit the lower envelope of the distribution in Figure \ref{fig-resck} as
\begin{equation}\label{eqn-rescklow}
\frac{S_i}{S_p} = \frac{1}{1 + (200/SNR^3)}
\end{equation}
where $S_i$ and $S_p$ represent the integrated and peak flux densities and $SNR$ is the signal-to-noise ratio, defined as the peak flux density divided by the local rms sensitivity. We determined the constant, 200, by finding its value such that 95\% of all sources with $S_i / S_p < 1$ are bound by the function. 

It is clear from Figure \ref{fig-resck} that the distribution of the ratio of peak to integrated flux density is not centered on unity. This is the aforementioned effect of combining separate pointings each with their own degrees of bandwidth smearing and the creation of the mosaic image not explicitly accounting for this bandwidth smearing. The resulting source will appear to have a core representing the region for which each pointing contributes to the flux density and a much fainter irregular extended halo consisting of the bandwidth-smeared flux densities of the separate pointings. The net effect is that sources with high signal-to-noise will always have an integrated flux density greater than their peak flux density, although due to the mosaic nature of the map such sources are not unambiguously resolved. 

We can model the expected size of this effect using reasonable assumptions. The spatial distribution of VLA antennas and our weighting of the resulting $(u,v,w)$ data imply that our $(u,v,w)$ coverage is approximately a circularly-symmetric Gaussian, and the individual 3.125~MHz channels after bandpass calibration can be assumed to have nearly square response. Under these simplifications, the reduction in measured amplitude of a point source caused by bandwidth smearing relative to its true amplitude is:
\begin{equation}\label{eqn-bwsmear}
\frac{I}{I_0} = \frac{\sqrt{\pi}}{\gamma \beta} \mbox{erf} \frac{\gamma \beta}{2}
\end{equation}
\citep{bridle1994}, where $I_0$ is the amplitude of the point source in the absence of bandwidth smearing, $\gamma \equiv 2 \sqrt{\ln 2}$, and $\beta = \frac{\Delta \nu}{\nu} \frac{\theta}{\theta_{FWHM}}$. This latter term is the fractional bandwidth of the observations (in our case, 3.125~MHz divided by 1.4~GHz) times the distance from the pointing center expressed in number of beams (the separation $\theta$ divided by the FWHM of the beam, $\theta_{FWHM}$). For simplicity in our estimation of the effect of bandwidth smearing we ``circularize'' the beam and assume it is symmetric with a FWHM of 2\farcs1. We can then apply Equation \ref{eqn-bwsmear} to determine the reduction in peak response associated with each pointing that contributes to the mosaic image at the position of each source, and weight these per-pointing responses in the same manner that was used to generate the mosaic image. This yields the reduction in peak response caused by bandwidth smearing, and since in the absence of noise the net flux density is conserved, the inverse of the left-hand side of Equation \ref{eqn-bwsmear} is the ratio of the integrated flux density to the peak flux density: the ratio examined in Figure \ref{fig-resck}. 

\begin{figure}
\figurenum{4}
\epsscale{0.9}
\plotone{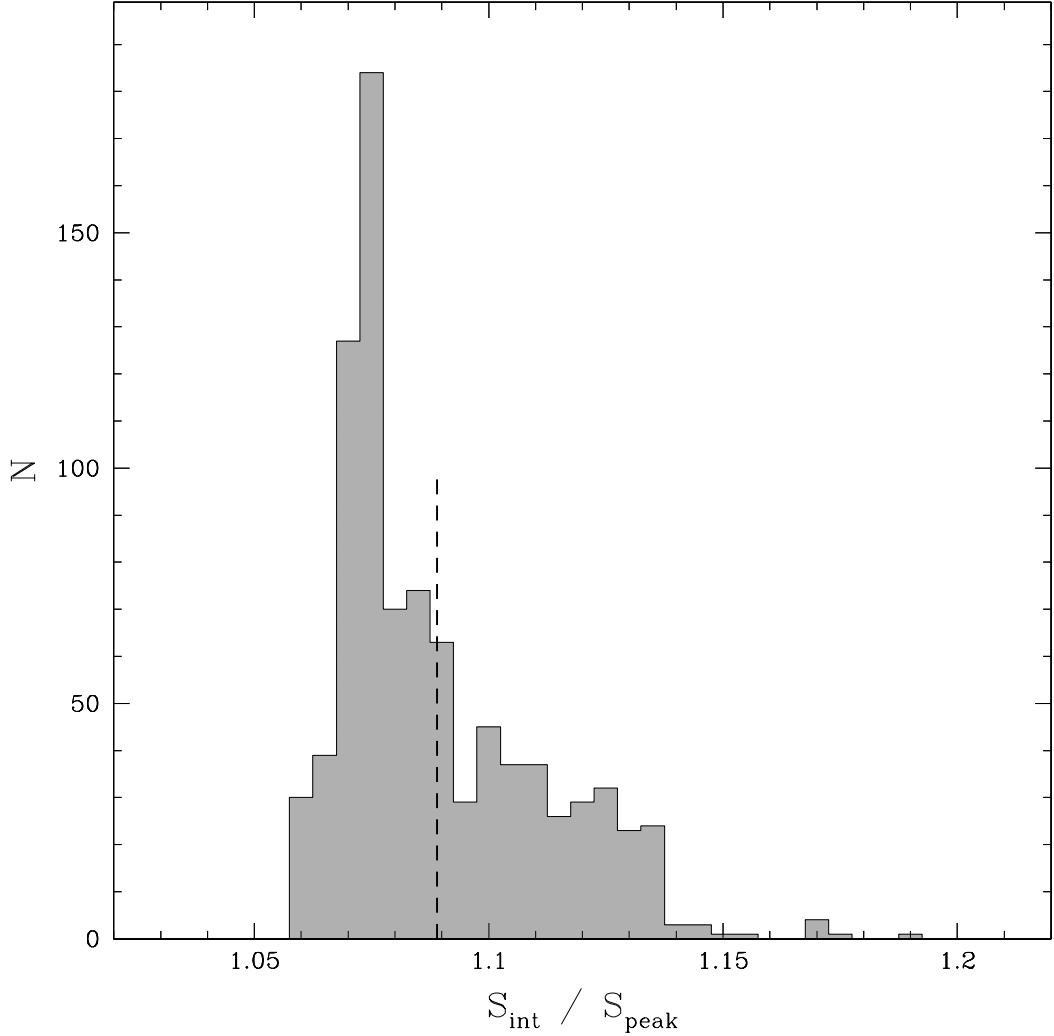}
\caption{Histogram of modeled bandwidth smearing at the positions of sources in the mosaic image. Bandwidth smearing reduces the peak flux density of sources but conserves the total flux density, so the ratio $S_i / S_p$ is greater than 1. The dashed vertical line marks the average, $S_i / S_p = 1.089$. By design of the survey, most sources lie reasonably close to a pointing center and thus the histogram peaks at a value of $S_i / S_p \sim 1.07$. Near the center of the pointing ring the sensitivity is best because all six pointings contribute to the mosaic image, but the distance to each pointing is relatively large so bandwidth smearing is greater. This causes the large tail to values of $S_i / S_p \lesssim 1.14$. The small number of sources with higher $S_i / S_p$ fall at the corners of the mosaic image.\label{fig-bwsmear}}
\end{figure}

We plot a histogram of the modeled effect of bandwidth smearing in Figure \ref{fig-bwsmear}, which is based on the positions of actual sources in our mosaic image. The results are consistent with the offset seen in Figure \ref{fig-resck}, with the mean ratio of $S_{i}$ to $S_{p}$ being $1.09$ with a dispersion of $0.02$. For perspective, the range in values is understood by considering limiting cases. The least possible bandwidth smearing occurs for a source coincident with a single pointing center, for which the mosaic map is dominated by data associated with that pointing but also includes contributions from its two adjacent pointings. In this case, the ratio of $S_{i}$ to $S_{p}$ is 1.06. At the other extreme, a source directly in the center of the mosaic image is equidistant from all six pointings and thus suffers modest bandwidth smearing from each. For such a source, the ratio of $S_{i}$ to $S_{p}$ is 1.14. It can be seen in Figure \ref{fig-bwsmear} that these values do indeed bracket the majority of sources and the peak in the histogram around $S_i / S_p \approx 1.07$ indicates that, by design, most sources fall reasonably close to a single pointing center. The small number of sources at $S_i / S_p > 1.14$ correspond to the corners of the mosaic image.

Returning to Figure \ref{fig-resck} and the evaluation of whether a source is resolved, instead of simply mirroring Equation \ref{eqn-rescklow} about $S_i / S_p = 1$ we include an offset to account for bandwidth smearing. Based on the spread of values for $S_i / S_p$ from bandwidth smearing shown in Figure \ref{fig-bwsmear} and its consistency with that measured for sources that appear unresolved based on the Gaussian source fitting to their single best pointing, we set the upper envelope for evaluating whether a source is resolved at
\begin{equation}\label{eqn-resckhigh}
\frac{S_i}{S_p} = 1.2 + (200/SNR^3)
\end{equation}
Sources for which $S_i / S_p$ is above this signal-to-noise based threshhold are thus considered to be resolved using this approach. This is indicated by a flag in the source catalog, and additional justification for the 20\% offset in the upper envelope is provided in Section \ref{sec-flux} when we compare our flux densities with those of single-pointing data.

The points in Figure \ref{fig-resck} are coded to reflect whether individual sources appear to be resolved under each of the two techniques. Grey asterisks represent all sources for which both fitted axes have minimum values of zero and the ratio of peak to integrated flux density lies below the curve described by Equation \ref{eqn-resckhigh}; these points were associated with unresolved sources in each technique. Similarly, open circles represent those sources that appear to be resolved when evaluated using both techniques. The filled black symbols represent the less conclusive cases, with filled triangles signifying sources that appear resolved in their best single-pointing data yet do not lie above the fitted relationship of Equation \ref{eqn-resckhigh} and filled circles signifying sources that appear unresolved in their single-pointing data yet lie above the fitted relationship thus suggesting that they are resolved.

\subsection{Multiple Component Sources}\label{sec-mult}

Powerful radio galaxies are often resolved into multiple components, with the classic example being FRII sources typically composed of a pair of bright, extended radio lobes surrounding the progenitor galaxy which sometimes is also coincident with a distinct radio core \citep{fanaroff1974}. The sensitivity and resolution of the survey can also ``break up'' the contiguous radio emission associated with a single progenitor galaxy into multiple apparently separate components. Possible associations are generally made by the experienced radio astronomers who collect, reduce, and analyze the radio data and are thus inherently subjective, although usually accurate. In M08, we relied upon the listing of identified multiple component sources from K08 and added new possible components that the slightly higher resolution of our survey provided.

In the present work, we take advantage of the outstanding available multiwavelength coverage to assess multiple component sources. \citet{bonzini2012} use a likelihood ratio technique to identify counterparts for the radio sources associated with this survey, relying on deep optical, near- and mid-infrared data. This is highly effective at assessing possible multiple component sources, as sometimes radio sources thought to be components associated with a nearby but spatially-separate galaxy are shown to have much stronger associations with a unique, fainter coincident galaxy. Similarly, the outstanding depth of the optical and IR data make radio sources without a coincident counterpart very rare. This means that close positional coincidence of a radio source with an optical or IR counterpart is almost certainly an association with a single galaxy rather than a component associated to a neighboring galaxy. Our radio catalog is thus guided by the counterpart identification. If a single galaxy has multiple associated radio components, we list their combined properties as a single entry in the main catalog and provide a flag indicating that the source consists of multiple components. A subsequent catalog then includes the information on each separate component of the combined source. There are 17 such multiple-component sources, with a total of 49 individual components ascribed to them. We stress that no information is lost by this approach; every distinct peak in the radio emission of the final image is included in one of the two catalogs, and users interested in the total radio emission associated with a given galaxy will find such information in the first catalog. Images of the 17 multiple-component sources are provided in the Appendix.

\subsection{Source Flux Density}\label{sec-flux}

Of greater practical importance to most users is the question of what flux density should be used for each source? As previously discussed, this is related to the issue of source morphology even if such detail is not needed by the user. In M08 we confirmed the general consistency of our flux density scale with prior radio observations of this region, particularly the deep single-pointing data of K08. We also confirmed that the flux densities based on Gaussian fitting were consistent with simple aperture photometry, indicating that although considerations such as bandwidth smearing do affect the apparent morphology of sources the standard approach of using Gaussian fits to evaluate flux densities was still appropriate. We rely on the same basic approach here to specify which flux density measurement, peak or integrated, should be adopted in order to most accurately reflect the true flux density of a source. This was accomplished by calculating the mean and dispersion of the ratio of flux density from our catalog with that of K08, using various prescriptions to choose between peak and integrated flux density measurements. In general, these prescriptions paralleled the discussion of source morphology and the two evaluations thereof, one based on whether the fitted deconvolved axial sizes were consistent with zero and one based on the ratio of integrated to peak flux density and the parametrized envelope bounding unresolved sources. These will henceforth be referred to as ``Fitted Axis'' and ``Envelope,'' respectively. For resolved sources we used the integrated flux density measurement while for unresolved sources we used the peak flux density measurement. We also explored the use of a threshold in $SNR$, using integrated flux densities for sources above the threshold and peak flux densities for those below. Combinations of these various prescriptions were also examined and Table \ref{tbl-flxck} summarizes the findings. 

The flux densities associated with the Fitted Axis method are slightly lower than those reported in K08 ($< S / S_{K08} >  = 0.985$), while the Envelope method yields slightly higher flux densities ($< S / S_{K08} >  = 1.020$). After performing a single round of 3$\sigma$ clipping to remove outliers (for example, sources with intrinsic variability over the $\sim$6 years between the K08 observations and this survey), the dispersions associated with the two methods are similar at $0.218$ and $0.225$, respectively. At this point we note that adding the 20\% offset to the upper curve in the Envelope method and using this to evaluate whether sources are resolved and hence whether to adopt their integrated flux density measurements has a large effect. Simply mirroring the envelope determined for $S_i / S_p < 1$ without including the offset produces flux densities that are about 7\% higher than those in K08. This is just a consequence of the convergence to $S_i / S_p \approx 1.1$ for brighter unresolved sources (see Figures \ref{fig-resck} and \ref{fig-bwsmear}). Since the two methods have opposing effects relative to the K08 flux density scale, we investigated logical combinations of ``OR'' (if either method indicated a source was resolved, its integrated flux density was adopted) and ``AND'' (a source was assumed to be resolved only if both methods agreed that it was). Neither of these produced an improvement on simply adopting one of the two methods, although the dispersion with the ``AND'' combination after clipping outliers was the lowest we found. 

These findings led us to identify a ``Hybrid'' solution. The ``Hybrid'' method applies a cut of $SNR = 20$, with all sources detected at greater than this threshold being represented by their integrated flux density measurement. For sources with lower $SNR$, the peak flux density measurement is adopted unless both the Fitted Axis and Envelope methods indicate the source is resolved in which case its integrated flux density is used. This prescription yields flux densities nearly identical to those of the K08 survey, $< S / S_{K08} >  = 1.005$ and $< S / S_{K08} >  = 0.997$ before and after clipping. The dispersion in the ratio of the flux densities after the clipping is also very low. We recommend this Hybrid solution for the selection of peak or integrated flux density, and indicate its selection in the final source table. 

\subsection{Final Catalog}

The final source catalog is presented in Tables \ref{tbl-cat} and \ref{tbl-multiple} and encapsulates all of the above discussion. Table \ref{tbl-cat} is the main catalog of radio sources, wherein the 17 sources thought to consist of multiple components associated with a single host object are listed with a single aggregate integrated flux density. Gaussian fits to the individual components associated with these sources are listed in full detail in Table \ref{tbl-multiple}. The data columns for each table are identical, and are summarized as follows:

(1) - (6) {\it Source position} (J2000). In most cases, the listed position is that of the center of the Gaussian that best fits the data. The typical position errors for strong, unresolved sources observed with the VLA at 1.4~GHz are 0\farcs1, and position errors attributed to the Gaussian fitting in the presence of noise are on the order of the beam size divided by twice the signal-to-noise ratio \citep{condon1997}. Thus, the position error in Right Ascension for a 5$\sigma$-point source would be $\sqrt{0.1^2 + \{1.6/(2 \times 5)\}^2} \approx 0\farcs2$ while the position error in declination for the same source would be $\sqrt{0.1^2 + \{2.8/(2 \times 5)\}^2} \approx 0\farcs3$. For some resolved sources, the radio emission is poorly fit by a Gaussian and the source is evaluated using an irregularly-shaped aperture that covers the apparent emission (AIPS task {\scshape tvstat}). These sources are flagged (column 29) and their positions are usually just the location of the maximum emission within the aperture.

(7) {\it Signal-to-noise ratio}. This is defined as the fitted peak flux density divided by the local rms noise (i.e., column 8 divided by column 9). 

(8) {\it Peak flux density} in units of $\mu$Jy per beam. 

(9) {\it Local rms noise} in units of $\mu$Jy per beam. This is evaluated using the RMS map and thus represents the local noise within a 135\arcsec{}-diameter circle around the source position.

(10) - (11) {\it Integrated flux density} and associated error, in units of $\mu$Jy. As with the prior columns, these values represent the Gaussian fit to the source unless otherwise noted. For the extended sources poorly fit by a Gaussian, the total flux density within the irregularly-shaped aperture is indicated and its error is just the square root of the number of beams covering the aperture times the local rms noise.

(12) - (14) {\it Source size and position angle from mosaic image}, in units of arcsec and degrees. These represent the fitted source size including convolution with the 2\farcs8 by 1\farcs6 beam, with the position angle measured in degrees east from north (e.g., the beam has $PA = 0$). In addition, these parameters include the differing contributions of bandwidth smearing from each of the pointings contributing to the mosaic image at the position of the source. Note that these columns are only present in the online version of the tables.

(15) {\it Best pointing}. The index number of the individual pointing which has the lowest rms noise at the position of the source. The pointings are numbered 1 through 6 starting due east of the center of the mosaic and progressing clockwise (see Table \ref{tbl-ptgs}).

(16) {\it rms noise in best pointing} in units of $\mu$Jy per beam. 

(17) - (21) {\it Source size and position angle from best pointing}, in units of arcsec and degrees. The sizes are presented as $\pm1\sigma$ range for the major axis (columns 17 and 18) and minor axis (columns 19 and 20), while the position angle (column 21) is just the nominal value. Since these values are based on Gaussian fitting to the source in the single pointing with the lowest rms, they have accounted for bandwidth smearing and have been deconvolved to remove the synthesized beam.

(22) {\it Extended flag}. If the source was found to be extended using the ``envelope'' method (see Section \ref{sec-morph}), this column is set to 1. Sources that were unresolved under this test have a value of 0 in this column.

(23) {\it Flux density choice flag}. The recommended flux density measurement to use for the source (see Section \ref{sec-flux}). If the value in this column is ``P'' the flux density is better represented by the peak flux density (column 8) whereas if it is ``I'' the integrated flux density (column 10) is recommended.

(24) {\it Pointings contributing to mosaic image}. This is a listing of the pointings that contribute to the mosaic image at the position of the source. Thus, if the source position falls within the $33\%$ power point of the primary beam (about 20\arcmin{}) associated with an individual pointing that pointing is noted here. This information allows users to inspect the images associated with the individual pointings to further assess source morphology and characteristics. This column is only available in the online version of the tables.

(25) - (28) {\it Source information from Kellermann et al. (2008)}. For sources that were identified in K08, the associated identification number (column 25), flux density and error (columns 26 and 27) in units of $\mu$Jy, and deconvolved source size (column 28, including upper limits) are provided. These columns are only available in the online version of the tables.

(29) {\it Notes}. Flags indicating details such as extended source morphologies and other source fitting details. Multiple-component sources are indicated here (Table \ref{tbl-cat}) and cross referenced to their individual fitted components (Table \ref{tbl-multiple}), with figures provided in the Appendix.

\section{Summary and Discussion}\label{sec-sum}

\begin{figure}
\figurenum{5}
\epsscale{0.9}
\plotone{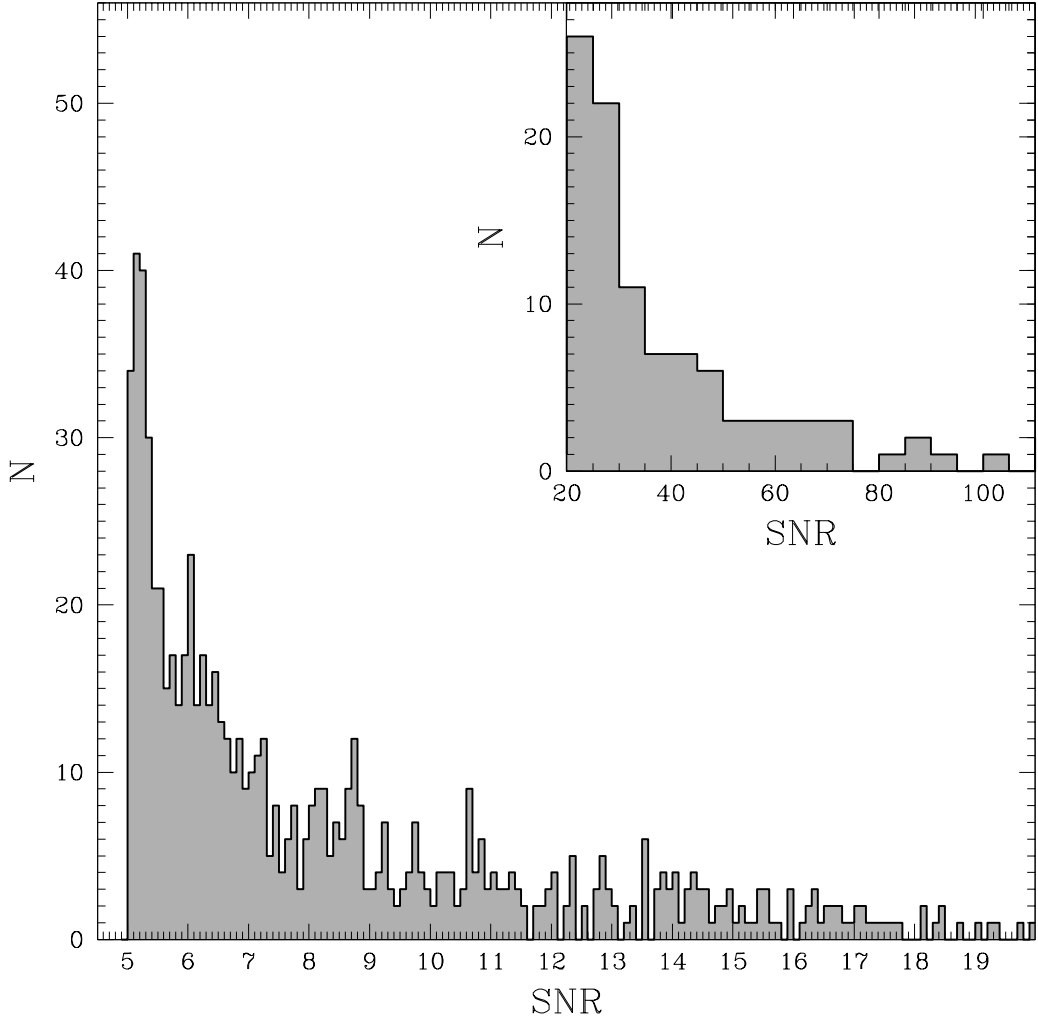}
\caption{Histogram of number of radio sources as a function of signal-to-noise ratio, created using all detected peaks in the mosaic image (i.e., inclusive of individual components of multiple-component sources). The main histogram uses a bin size of 0.1, and the inset for higher signal-to-noise ratios uses a bin size of 5. There are an additional 47 sources with $SNR > 100$.\label{fig-snrhist}}
\end{figure}

Table \ref{tbl-cat} provides a listing of 883 distinct radio sources, 17 of which are single entries for multiple-component sources where the total radio emission appears to be associated with a single host galaxy. The details on the individual separate components for these 17 sources form the basis for the second table, which contains a total of 49 separate components for these sources. To summarize, the final mosaic image consists of 915 separate peaks in radio emission (883 in Table \ref{tbl-cat} minus the 17 multiple component sources, plus the 49 individual components of these sources listed in Table \ref{tbl-multiple}). For comparison, the early data release catalog of M08 consisted of 495 unique components. Figure \ref{fig-snrhist} reveals the primary reason for this greatly increased number, as the $SNR$ limit for inclusion has decreased from 7 to 5 for this catalog relative to the first data release. In addition, the $SNR$ for any given source will usually be increased in the present catalog on account of the improved imaging, which typically amounts to an rms noise that is 0.5 $\mu$Jy lower than that of the first data release at any point in the mosaic image.

In addition to the catalogs, this second data release consists of 9 images and 6 calibrated $(u,v,w)$ datasets. These all may be accessed online\footnote{http://www.astro.umd.edu/{\fsym \symbol{24}}nmiller/VLA\_ECDFS.html}. There are individual images and calibrated $(u,v,w)$ files for each of the six pointing centers (refer to Table \ref{tbl-ptgs}), along with the main mosaic image that uses all six pointings and its associated RMS image. The final released image is a larger area mosaic that incorporates all data out to the $33\%$ power point of the VLA primary beam. While there is no catalog associated with the additional area provided by this larger mosaic, it is useful for identification and measurement of sources found in other wide-area surveys.

\begin{figure}
\figurenum{2}
\epsscale{0.9}
\plotone{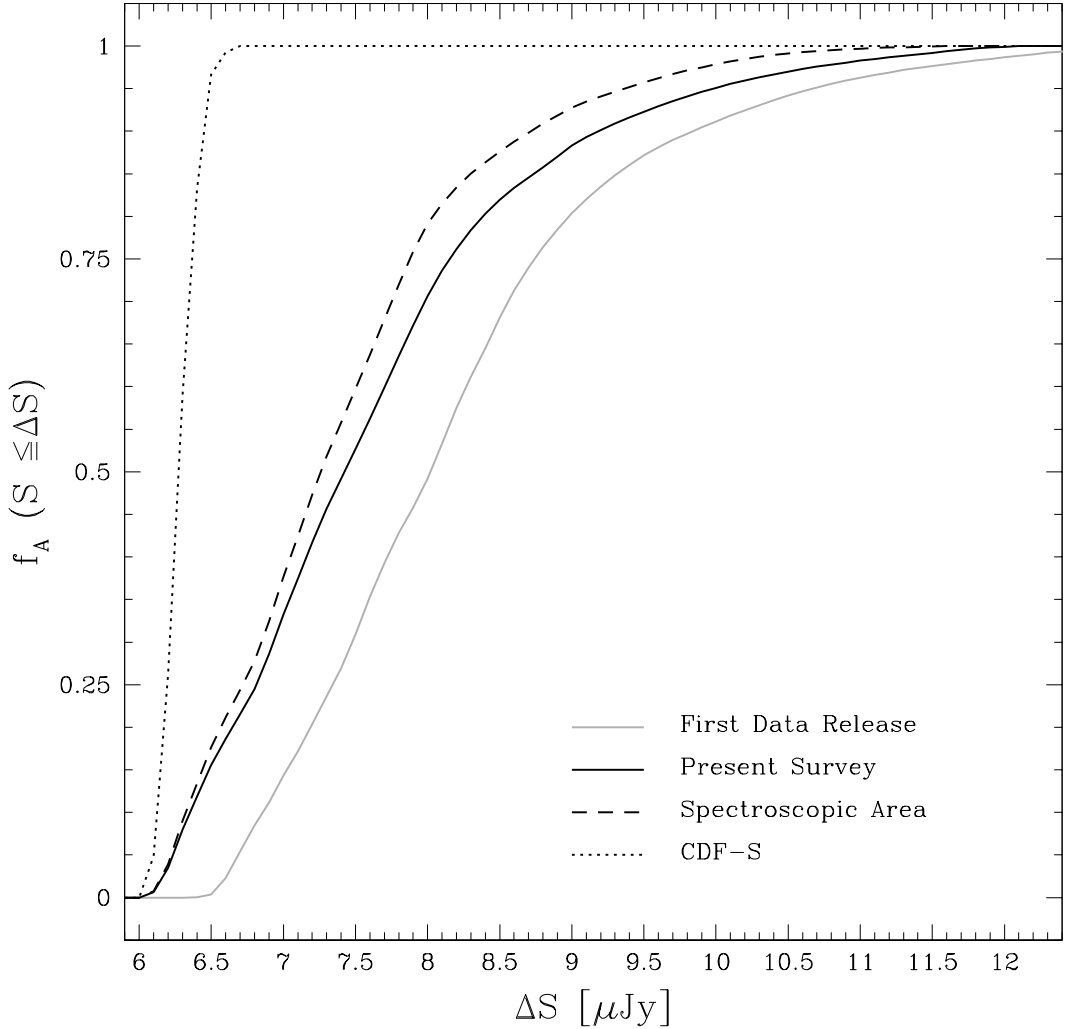}
\caption{Fractional area covered at a given sensitivity or better, for the released mosaic image of 34\farcm1 by 34\farcm1. The image described in this paper is shown in black, while the results presented in the first data release are shown in grey. The improvement is about 0.5$\mu$Jy across the entire area. Also shown are the fractional areas covered at a given sensitivity or better for two smaller regions within the mosaic image, corresponding to a large region with multiwavelength data for quality photometric redshifts (dashed line) and a small central region corresponding to the deepest uniform coverage within the CDF-S 4~Msec data (dotted line).\label{fig-rms}}
\end{figure}

The depth of the survey makes the surface density of radio sources competitive with that determined in X-rays. For the full mosaic image (0.324 square degree) and its associated 5$\sigma$ catalog presented in Table \ref{tbl-cat}, there are over 2,700 sources per square degree. This is over 30 times the source density associated to the Faint Images of the Radio Sky at Twenty centimeters survey \citep[FIRST;][]{white1997}. Because the sensitivity is fairly uniform across the mosaic image this figure increases only slightly towards the image center. For the portion of the mosaic image with good multiwavelength coverage to provide photometric redshifts (refer to Section \ref{sec-image:sum} and Figure \ref{fig-rms}) the source density approaches 2,800 sources per square degree, and for the region of deep and uniform coverage from the CDF-S 4~Msec observations (i.e., the 7\arcmin{} radius about the exposure-weighted center point) it is nearly 3,100 sources per square degree. In the central 5\arcmin{} around the center of the mosaic (i.e., the center of the pointing ring) where the rms sensitivity ranges from 6 $\mu$Jy per beam to 6.6 $\mu$Jy per beam, the density is over 3,500 sources per square degree. For comparison, the X-ray source density within the inner 3\arcmin{} of the 4~Msec CDF-S is roughly 16,700 sources per square degree \citep{xue2011} but over the full CDF-S coverage it is only about twice the surface density of radio sources although the observing time for the radio observations was about a factor of six less than for the X-ray observations.


Similarly, simple positional matching of X-ray and radio catalogs provides some illuminating numbers. Considering just the region of deep and uniform coverage from the CDF-S 4~Msec observations, 16.8\% (80/475) of the \citet{xue2011} X-ray sources have matches in the radio catalog using a simple positional match of 2\arcsec . The detection fraction within this same area only rises slightly as the {\it Chandra} integration time declines, being 18.2\% (50/274) for the CDF-S 2~Msec catalog of \citet{luo2008}, 19.0\% (38/200) for the CDF-S 1~Msec catalog of \citet{giacconi2002}, and 21.8\% (19/87) for the shallower E-CDF-S catalog of \citet{lehmer2005}. Looking at this from the other direction, the increasing integrations with {\it Chandra} are gradually producing larger detection fractions of the radio sources. While only 14.5\% of the radio sources within this region are detected in the E-CDF-S observations (19/131), 29.0\% (38/131) are detected in the 1~Msec data, 38.2\% (50/131) are detected in the 2~Msec data, and 61.1\% (80/131) are detected in the full 4~Msec data. At this rate of increase, should the CDF-S be expanded to 10~Msec \citep[e.g.,][]{lehmer2012} we can expect X-ray detections for over 90\% of the radio sources from within this central area.


Emphasis on the E-CDF-S continues with additional radio surveys either in process or planned. The rebirth of the VLA as the Karl G.\ Jansky Very Large Array, with new receivers and electronics as well as a wide-bandwidth, many channel correlator, will greatly advance this objective. The RMS sensitivity of a radio image is inversely proportional to the square root of the bandwidth, meaning the new correlator produces a dramatic improvement in the depth that VLA images can achieve in reasonable duration programs. Consequently, future deep fields will routinely achieve RMS sensitivities of $\sim1$ $\mu$Jy at frequencies near 1.4~GHz \citep[e.g.,][]{condon2012}. This will translate to thousands of detected radio sources in the E-CDF-S, and the wide bandwidth will also provide spectral indices for brighter sources where signal-to-noise is high. One of the tradeoffs in survey imaging is the size of the primary beam (and hence field of view), which is inversely proportional to the frequency of observation. This means that doubling the frequency of observation requires four times as many pointings to cover the same area, and the attendant increase in program duration has greatly limited the number of wide-field surveys performed at frequencies above 1.4~GHz. The vast improvement in sensitivity arising from increased bandwidth will counteract this problem, and deep surveys at higher observing frequencies will also be achievable in reasonable observing times. These will provide spectral measurements for a large fraction of the sources. 

\acknowledgements
NAM gratefully acknowledges support through {\it Chandra} Award AR8-9016X, and a Jansky Fellowship of the National Radio Astronomy Observatory held during the period when these data were obtained.

\begin{deluxetable}{l l l r}
\tablecolumns{4}
\tablecaption{Pointing Coordinates\label{tbl-ptgs}}
\tablewidth{0pt}
\tablehead{
\colhead{Pointing ID} & \colhead{RA(J2000)} & \colhead{Dec(J2000)} & \colhead{RMS\tablenotemark{a}}}
\startdata
ECDFS~1\tablenotemark{b}  & 03:33:22.25 & -27:48:30.0 & 10.5~$\mu$Jy \\
ECDFS~2  & 03:32:55.12 & -27:38:03.0 & 9.4~$\mu$Jy \\
ECDFS~3  & 03:32:00.88 & -27:38:03.0 & 9.7~$\mu$Jy \\
ECDFS~4  & 03:31:33.75 & -27:48:30.0 & 9.5~$\mu$Jy \\
ECDFS~5  & 03:32:00.88 & -27:58:57.0 & 10.0~$\mu$Jy \\
ECDFS~6  & 03:32:55.12 & -27:58:57.0 & 9.3~$\mu$Jy \\
\enddata

\tablenotetext{a}{RMS sensitivity for final image associated with all data for that pointing, prior to correction for primary beam.}
\tablenotetext{b}{One observation intended for this position was erroneously offset to 03:33:22.25 -27:47:30.0. This offset observation is not included in the pointing 1 data imaged to produce the indicated RMS sensitivity.}

\end{deluxetable}

\begin{deluxetable}{l r r r r r}
\tablecolumns{6}
\tabletypesize{\small}
\tablecaption{Flux Density Comparisons\label{tbl-flxck}}
\tablewidth{0pt}
\tablehead{
\colhead{} & \multicolumn{2}{c}{Full Sample} & \multicolumn{3}{c}{Clipped Sample} \\
\colhead{Method} & \colhead{Mean} & \colhead{Disp} &
\colhead{Mean} & \colhead{Disp} & \colhead{N$_{clip}$}
}
\startdata
Fitted Axis & 0.985 & 0.248 & 0.978 & 0.218 &  6 \\
Envelope    & 1.085 & 0.290 & 1.071 & 0.256 &  6 \\
Envelope+20 & 1.020 & 0.276 & 0.994 & 0.225 & 10 \\
Both-OR     & 1.041 & 0.278 & 1.013 & 0.227 & 11 \\
Both-AND    & 0.964 & 0.241 & 0.960 & 0.213 &  5 \\
SNR         & 0.938 & 0.421 & 0.938 & 0.421 &  0 \\
Hybrid      & 1.005 & 0.258 & 0.997 & 0.220 &  6 \\
\enddata

\tablecomments{{\footnotesize Columns: (1) Method used to select between peak and integrated flux density measurement. For each method, sources considered to be resolved are described by their integrated flux density while unresolved sources are described by their peak flux density. In ``Fitted Axis,'' sources for which the fitted minimum of the major axis is greater than zero are considered resolved; in ``Envelope,'' sources with integrated to peak flux density ratios above the envelope described by the reflection of Equation \ref{eqn-rescklow} above integrated equals peak flux density are considered resolved; in ``Envelope+20,''  sources with integrated to peak flux density ratios above the envelope described by Equation \ref{eqn-resckhigh} (i.e., the reflected envelope of Equation \ref{eqn-rescklow} offset upward by 20\%) are considered resolved; in ``Both-OR'' and ``Both-AND'' the two methods are combined with logical operators, so that a source is considered resolved if either method indicates that it is (``OR'') or only if both methods indicate that it is (``AND''); in ``SNR'' sources with $SNR>10$ have their integrated flux densities adopted, while sources below this threshhold use peak flux densities; and in ``Hybrid'' a mix of the methods is adopted, such that for sources with $SNR>20$ the integrated flux density is used while for sources below this threshhold the peak flux density is used unless both methods indicate a resolved source, in which case the integrated flux density is used. Numbers in Columns (2) through (5) refer to the statistics based on the ratio of flux density from the current catalog to that of the K08 catalog, with (2) and (3) being the mean and dispersion based on all common objects and (4) and (5) being the mean and dispersion after clipping of sources that differ from the mean by more than 3$\sigma$. Column (6) indicates the number of outlier objects removed by this single iteration of clipping.}}

\end{deluxetable}

\begin{deluxetable}{r r r r r r r r r r r r r r r r r r r r r r r r r r r r r}
\tablecolumns{21}
\tabletypesize{\tiny}
\rotate
\tablecaption{Main Catalog of Radio Sources\label{tbl-cat}}
\tablewidth{0pt}
\tablehead{
\multicolumn{3}{c}{RA} & \multicolumn{3}{c}{Dec} & \colhead{$SNR$} & \colhead{$S_p$} &
\colhead{RMS} & \colhead{$S_i$} & \colhead{$e\_ S_i$} &
\colhead{Best} & \colhead{RMS$_{best}$} &
\multicolumn{2}{c}{Maj$_{best}$} & \multicolumn{2}{c}{Min$_{best}$} & \colhead{PA$_{best}$} &
\colhead{Ext?} & \colhead{Choice} & \colhead{Notes} \\
\multicolumn{3}{c}{(J2000)} & \multicolumn{3}{c}{(J2000)} & \colhead{} & \colhead{[$\mu$Jy~bm$^{-1}$]} &
\colhead{[$\mu$Jy~bm$^{-1}$]} & \colhead{[$\mu$Jy]} &  \colhead{[$\mu$Jy]} &
\colhead{Ptg} &  \colhead{[$\mu$Jy~bm$^{-1}$]} &
\multicolumn{2}{c}{[\arcsec ]} & \multicolumn{2}{c}{[\arcsec ]} & \colhead{[$^\circ$]} &
\colhead{} & \colhead{} & \colhead{} \\
\colhead{(1)} & \colhead{(2)} & \colhead{(3)} & \colhead{(4)} & \colhead{(5)} & \colhead{(6)} & 
\colhead{(7)} & \colhead{(8)} & \colhead{(9)} & \colhead{(10)} & \colhead{(11)} & 
\colhead{(15)} & \colhead{(16)} & \colhead{(17)} & \colhead{(18)} & 
\colhead{(19)} & \colhead{(20)} & \colhead{(21)} & \colhead{(22)} & \colhead{(23)} &
\colhead{(29)} 
}
\startdata
 3 & 31 & 10.69 & -28 & 03 & 22.8 &   18.7 &  207.5 & 10.6 &   365.2 & 26.9 & 5 & 14.6 &  1.4 & 2.3 & 0.0 & 0.7 & 169 & 1 & I & \nodata \\
 3 & 31 & 10.81 & -27 & 55 & 52.8 &   13.8 &  123.1 &  8.8 &   156.6 & 18.0 & 4 & 11.8 &  0.0 & 2.0 & 0.0 & 1.4 & 120 & 0 & P & \nodata \\
 3 & 31 & 10.83 & -27 & 55 & 57.8 &    5.3 &   47.3 &  8.9 &    32.7 & 12.1 & 4 & 11.9 &  0.0 & 3.1 & 0.0 & 0.0 &   2 & 0 & P & \nodata \\
 3 & 31 & 10.93 & -27 & 49 & 55.5 &    5.1 &   40.8 &  8.0 &    35.3 & 12.6 & 4 & 10.2 &  0.0 & 2.6 & 0.0 & 2.7 & 168 & 0 & P & \nodata \\
 3 & 31 & 10.95 & -27 & 58 & 10.4 &   12.7 &  121.2 &  9.5 &   136.9 & 17.9 & 4 & 12.9 &  0.0 & 1.3 & 0.0 & 1.5 &  21 & 0 & P & \nodata \\
 3 & 31 & 11.48 & -27 & 52 & 59.0 &   12.8 &  105.9 &  8.0 &   185.8 & 20.5 & 4 & 10.9 &  0.0 & 2.7 & 0.0 & 1.9 & 165 & 1 & P & \nodata \\
 3 & 31 & 11.63 & -27 & 55 & 06.3 &    5.4 &   47.1 &  8.7 &    69.0 & 20.1 & 4 & 11.4 &  0.0 & 4.7 & 0.0 & 1.6 &  49 & 0 & P & \nodata \\
 3 & 31 & 11.69 & -27 & 31 & 44.2 &  211.2 & 2555.\phn & 12.1 & 3232.\phn & 65.4 & 3 & 15.2 & 0.2 & 0.5 & 0.0 & 0.0 & 29 & 1 & I &    a \\
 3 & 31 & 11.84 & -27 & 58 & 18.0 &    8.1 &   77.2 &  9.4 &    92.5 & 18.4 & 4 & 12.7 &  0.0 & 2.9 & 0.0 & 1.3 & 162 & 0 & P & \nodata \\
 3 & 31 & 11.89 & -27 & 59 & 52.3 &    6.8 &   65.8 &  9.5 &    90.3 & 20.5 & 5 & 13.7 &  0.0 & 2.7 & 0.0 & 0.6 &  22 & 0 & P & \nodata \\
\enddata

\tablecomments{Only ten rows of data are displayed here to demonstrate the format, and columns (12) - (14) and (24) - (28) are omitted in this print version. The full table is available in the electronic version of the journal. See text for description of columns.}

\end{deluxetable}

\begin{deluxetable}{r r r r r r r r r r r r r r r r r r r r r r r r r r r r r}
\tablecolumns{21}
\tabletypesize{\tiny}
\rotate
\tablecaption{Individual Components of Multiple-Component Radio Sources\label{tbl-multiple}}
\tablewidth{0pt}
\tablehead{
\multicolumn{3}{c}{RA} & \multicolumn{3}{c}{Dec} & \colhead{$SNR$} & \colhead{$S_p$} &
\colhead{RMS} & \colhead{$S_i$} & \colhead{$e\_ S_i$} &
\colhead{Best} & \colhead{RMS$_{best}$} &
\multicolumn{2}{c}{Maj$_{best}$} & \multicolumn{2}{c}{Min$_{best}$} & \colhead{PA$_{best}$} &
\colhead{Ext?} & \colhead{Choice} & \colhead{Notes} \\
\multicolumn{3}{c}{(J2000)} & \multicolumn{3}{c}{(J2000)} & \colhead{} & \colhead{[$\mu$Jy~bm$^{-1}$]} &
\colhead{[$\mu$Jy~bm$^{-1}$]} & \colhead{[$\mu$Jy]} &  \colhead{[$\mu$Jy]} &
\colhead{Ptg} &  \colhead{[$\mu$Jy~bm$^{-1}$]} &
\multicolumn{2}{c}{[\arcsec ]} & \multicolumn{2}{c}{[\arcsec ]} & \colhead{[$^\circ$]} &
\colhead{} & \colhead{} & \colhead{} \\
\colhead{(1)} & \colhead{(2)} & \colhead{(3)} & \colhead{(4)} & \colhead{(5)} & \colhead{(6)} & 
\colhead{(7)} & \colhead{(8)} & \colhead{(9)} & \colhead{(10)} & \colhead{(11)} & 
\colhead{(15)} & \colhead{(16)} & \colhead{(17)} & \colhead{(18)} & 
\colhead{(19)} & \colhead{(20)} & \colhead{(21)} & \colhead{(22)} & \colhead{(23)} &
\colhead{(29)} 
}
\startdata
 3 & 31 & 13.99 & -27 & 55 & 19.9 &  144.4 & 1242.\phn & 8.6 & 7180.\phn &  57.3 & 4 & 11.3 &  7.1 &  7.2 & 2.0 & 2.1 &  81 & 1 & P & 1 \\
 3 & 31 & 15.06 & -27 & 55 & 18.9 &  120.9 & 1040.\phn & 8.6 & 1206.\phn &  16.5 & 4 & 11.2 &  0.0 &  0.6 & 0.0 & 0.7 &  42 & 1 & P & 1 \\
 3 & 31 & 17.05 & -27 & 55 & 15.  &  103.6 &  870.0    & 8.4 & 3650.\phn &  43.0 & 4 & 11.0 &  4.6 &  4.7 & 3.0 & 3.1 & 157 & 1 & P & 1 \\
\hline
 3 & 31 & 14.69 & -28 & 01 & 51.7 &    8.7 &   90.1   & 10.4 & 1367.\phn & 168.5 & 5 & 13.6 & 16.0 & 20.9 & 1.2 & 2.8 &  77 & 1 & P & 2 \\
 3 & 31 & 17.35 & -28 & 01 & 47.4 &   26.5 &  270.2 & 10.2 &  358.8      &  21.0 & 5 & 12.9 &  0.0 &  1.6 & 0.0 & 1.0 &  48 & 1 & P & 2 \\
 3 & 31 & 20.16 & -28 & 01 & 46.2 &    9.9 &   98.5 &  9.9 &  856.7      &  95.5 & 5 & 12.4 &  5.1 &  6.7 & 3.1 & 4.6 & 100 & 1 & P & 2 \\
\hline
 3 & 31 & 24.83 & -27 & 52 & 08.9 & 1043.\phn & 8032.\phn & 7.7 & 24670.\phn & 30.5 & 4 & 10.2 & 4.0 & 4.1 & 1.7 & 1.7 & 64 & 1 & P & 3 \\
 3 & 31 & 25.01 & -27 & 52 & 07.7 &  960.5 & 7396.\phn & 7.7 & 15380.\phn & 22.3 & 4 & 10.2 &  2.1 &  2.2 & 2.0 & 2.0 & 179 & 1 & P & 3 \\
\enddata

\tablecomments{Only eight rows of data are displayed here to demonstrate the format, and columns (12) - (14) and (24) - (28) are omitted in this print version. The full table is available in the electronic version of the journal. See text for description of columns.}

\end{deluxetable}

\clearpage

\appendix

\section{Multiple-Component Source Images}

Figures for each of the 17 multiple-component sources described in Section \ref{sec-mult} and listed in Table \ref{tbl-multiple} are presented here. Unless otherwise noted, each image is 1\arcmin{} by 1\arcmin{} with the greyscale ranging from $-35$ $\mu$Jy~beam$^{-1}$ to 105 $\mu$Jy~beam$^{-1}$. Contours are depicted at 5, 8, 13, 21, ... , and 987 times the local RMS noise. Information regarding distinct sources may be found in \citet{bonzini2012}.

\begin{figure}
\figurenum{A.1}
\epsscale{0.4}
\plotone{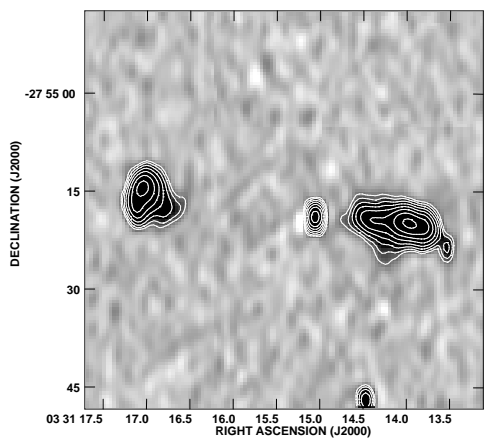}
\caption{Multiple-component source at 033115.0-275519, consisting of three components (core, east and west lobes). Note that the point source just to the edge of the western lobe (033113.53-275524.3) is a separate source associated with a counterpart in \citet{bonzini2012}.}
\end{figure}

\begin{figure}
\figurenum{A.2}
\epsscale{0.4}
\plotone{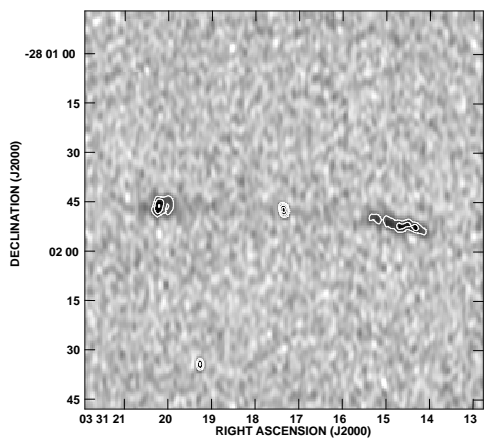}
\caption{Multiple-component source at 033117.3-280147, consisting of three components (core, east and west lobes). The size of this image is 2\arcmin{} by 2\arcmin .}
\end{figure}

\newpage

\begin{figure}
\figurenum{A.3}
\epsscale{0.4}
\plotone{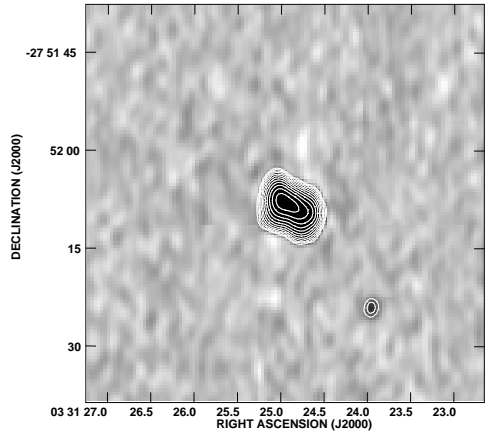}
\caption{Multiple-component source at 033124.9-275208, consisting of two blended components.}
\end{figure}

\begin{figure}
\figurenum{A.4}
\epsscale{0.4}
\plotone{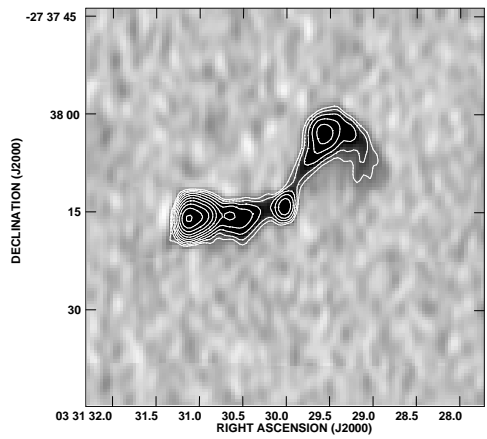}
\caption{Multiple-component source at 033130.0-273814, consisting of four components. These correspond to the core, the extended lobe to the northwest, and a pair of blended components comprising the eastern lobe.}
\end{figure}

\newpage

\begin{figure}
\figurenum{A.5}
\epsscale{0.4}
\plotone{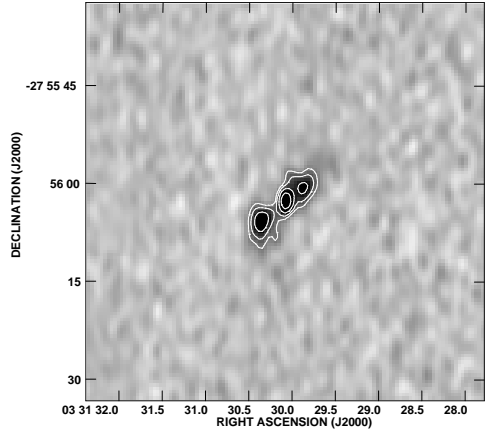}
\caption{Multiple-component source at 033130.1-275603, consisting of three components (core, northwest and southeast lobes).}
\end{figure}

\begin{figure}
\figurenum{A.6}
\epsscale{0.4}
\plotone{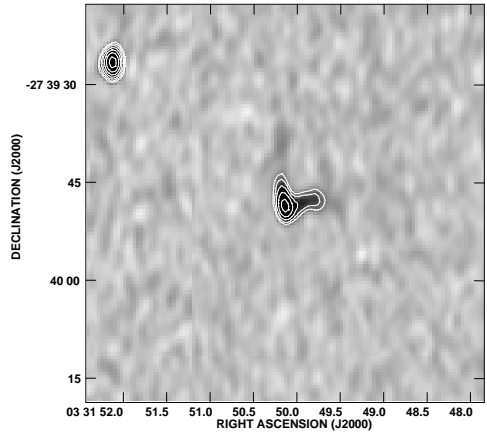}
\caption{Multiple-component source at 033150.1-273948, consisting of three components. The source has been deblended into a main component and an extension to the north, along with a third component extending to the west. The bright source at the northeast corner of the image is unrelated.}
\end{figure}

\newpage

\begin{figure}
\figurenum{A.7}
\epsscale{0.4}
\plotone{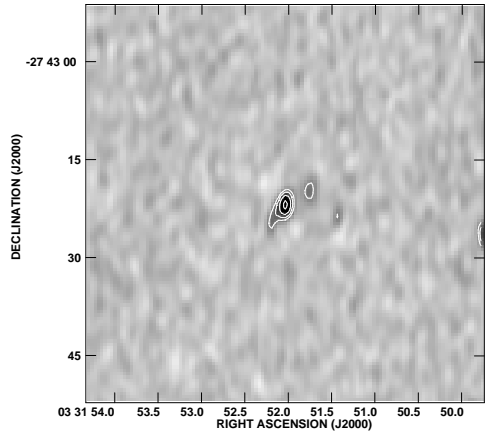}
\caption{Multiple-component source at 033152.0-274322, consisting of three components. The core has been deblended into a main core and an extension to the southeast, with the third component being the faint emission to the northwest.}
\end{figure}

\begin{figure}
\figurenum{A.8}
\epsscale{0.4}
\plotone{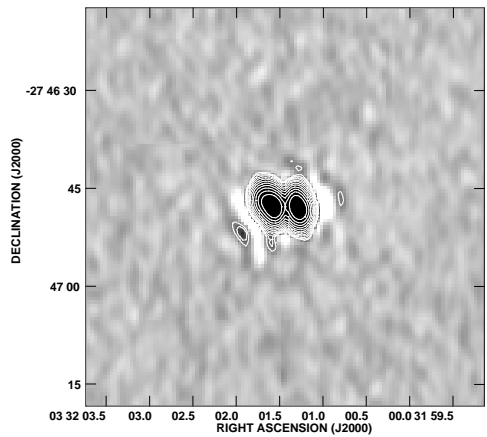}
\caption{Multiple-component source at 033201.4-274648, consisting of a compact double with overlapping lobes to the east and west.}
\end{figure}

\newpage

\begin{figure}
\figurenum{A.9}
\epsscale{0.4}
\plotone{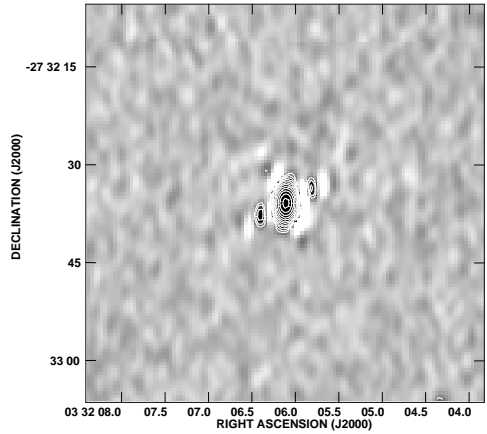}
\caption{Multiple-component source at 033206.1-273236. This is almost certainly a single bright ($>11$ mJy) source with sidelobes. The decreased dynamic range of the image is caused by this bright source lying near the northern edge of the mosaic image.}
\end{figure}

\begin{figure}
\figurenum{A.10}
\epsscale{0.4}
\plotone{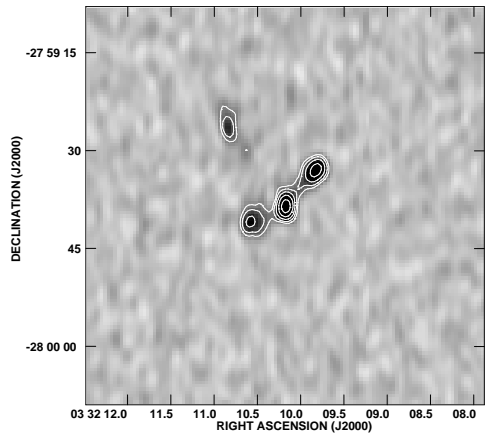}
\caption{Multiple-component source at 033210.2-275938, consisting of three components (core plus northwest and southeast lobes).}
\end{figure}

\newpage

\begin{figure}
\figurenum{A.11}
\epsscale{0.4}
\plotone{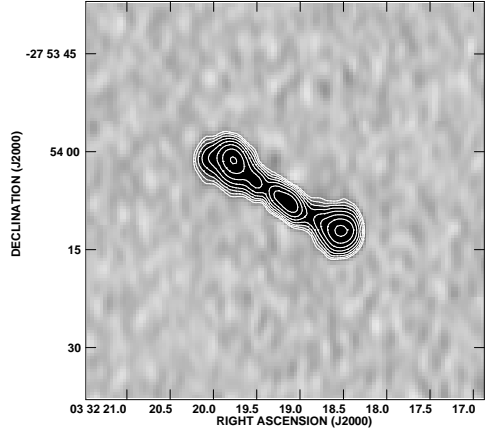}
\caption{Multiple-component source at 033219.2-275407, consisting of three components (core plus northeast and southwest lobes).}
\end{figure}

\begin{figure}
\figurenum{A.12}
\epsscale{0.4}
\plotone{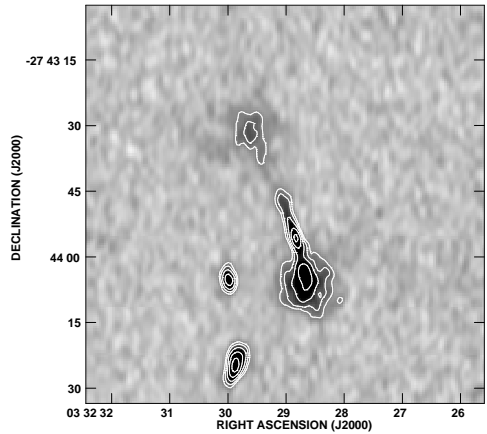}
\caption{Multiple-component source at 033228.8-274356, consisting of four components. This image is 1\farcm5 by 1\farcm5{}, and consists of a core, a component associated with the northern jet, the diffuse northern lobe, and the brighter southern lobe. Note that the pair of bright sources to the east and south are individual sources with counterpart identifications \citep{bonzini2012} and not components of this source.}
\end{figure}

\newpage

\begin{figure}
\figurenum{A.13}
\epsscale{0.4}
\plotone{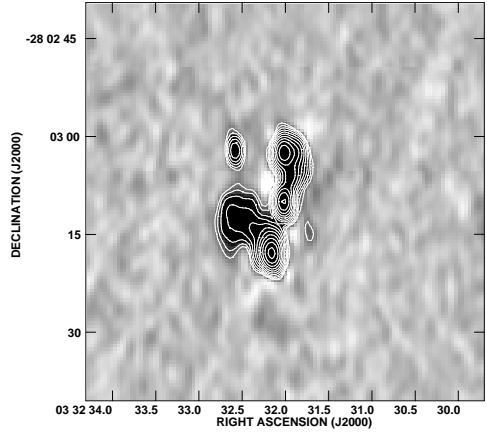}
\caption{Multiple-component source at 033232.0-280310, consisting of four components. These are the core, the lobe to the north, and the lobe to the south deblended from its diffuse emission extending toward the east. The bright source just to the northeast is associated with a unique counterpart \citep{bonzini2012} and is thus unrelated to the multiple-component source.}
\end{figure}

\begin{figure}
\figurenum{A.14}
\epsscale{0.4}
\plotone{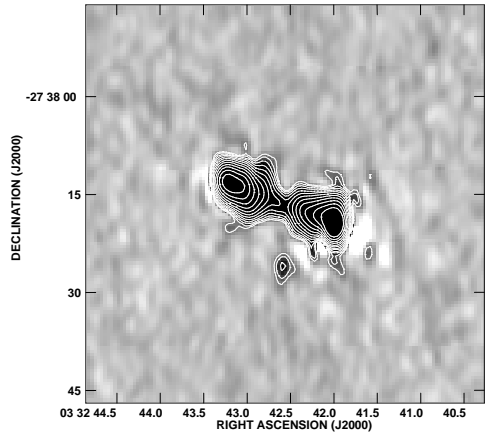}
\caption{Multiple-component source at 033242.6-273816. This is the brightest radio source in the field and it consists of a pair of bright radio lobes. The apparent low-level structures at the edges of the source are imaging artifacts. The point source just to the south is associated with a separate galaxy, and is thus unassociated with this complex.}
\end{figure}

\newpage

\begin{figure}
\figurenum{A.15}
\epsscale{0.4}
\plotone{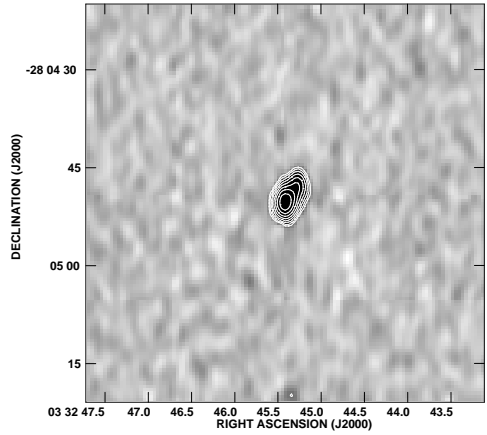}
\caption{Multiple-component source at 033245.4-280450, which has been deblended into a pair of components (core and extension to the northwest).}
\end{figure}

\begin{figure}
\figurenum{A.16}
\epsscale{0.4}
\plotone{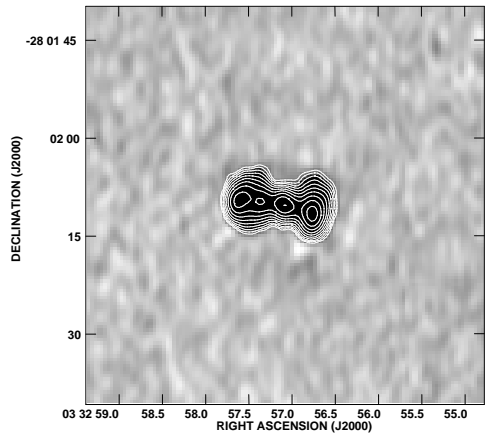}
\caption{Multiple-component source at 033257.1-280210, consisting of four components. These are a core, a western lobe, and the deblending of the eastern lobe into a pair of components.}
\end{figure}

\newpage

\begin{figure}
\figurenum{A.17}
\epsscale{0.4}
\plotone{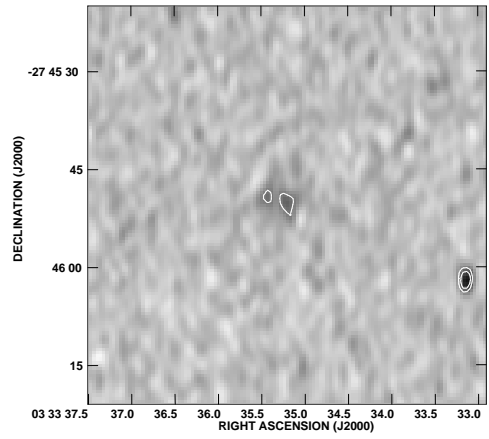}
\caption{Multiple-component source at 033335.2-274549, which has been deblended into a pair of faint components. These are both associated with disk emission in a spiral galaxy \citep{bonzini2012}.}
\end{figure}

\end{document}